# Isostructural Softening of Vulcanized Nanocomposites

Guilhem P. Baeza[1,*], Florent Dalmas[1], Fabien Dutertre[2] and Jean-Charles Majesté[2]

guilhem.baeza@insa-lyon.fr

[1] Univ Lyon, INSA-Lyon, CNRS, MATEIS, UMR5510, F-69621, Villeurbanne, France

[2] Univ Lyon, UJM-Saint-Etienne, CNRS, IMP, UMR 5223, F-42023, Saint-Etienne, France

**Abstract:**

Following a previous work evidencing that short poly-propylene glycol (PPG) chains incorporated to crude SBR/silica nanocomposites act as filler-network softeners without changing their structure, we propose in the present letter to examine more operative cross-linked materials. We first evidence that the adsorption of PPG onto silica deactivates progressively the particle's catalytic effect on vulcanization, without perturbing however the cross-link density distribution that we investigate through multiple-quantum NMR. In addition, electron microscopy confirms that silica structure is conserved after vulcanization and that it does not depend on the PPG content either. Composites containing various amount of PPG can thus be seen as structurally identical, both from a matrix and filler point of view – which is confirmed by small and medium amplitude oscillation shear rheology showing strikingly identical viscoelastic properties. PPG signature only appears above 100% in tensile deformation where it is seen to soften dramatically the filler network. Our discovery makes it consequently possible to decorrelate the mechanical behavior of reinforced rubbers under normal conditions of use and urgent needs of energy dissipation.

Nanocomposites based on statistical styrene-butadiene copolymer (SBR) reinforced with fractal aggregates of nano-silica have represented a breakthrough for tires manufacturers in the late 1980's by enabling the limitation of rolling resistance while keeping other performances satisfying.[1] This famous example, among others, lies in the decorrelation of mechanical properties, which opens perspectives far beyond the usual "compromise" philosophy mostly driving materials formulation. In the past, we have notably evidenced in such nanocomposites, the possible decorrelation of the plateau ($G' = G_N$) and loss ($G''$) moduli in a frequency (or temperature) window of interest.[2] This was achieved by tuning the crude materials composition through two parameters: (i) the fraction of graftable SBR, and (2) the molecular weight of the chains. In this way, it was possible to produce "twin" composites characterized by the same absolute grafting density (chains/silica nm²) while containing chains of different molecular weights. Strikingly, the grafting density was identified to be the unique structure-determining parameter, controlling in addition, the plateau modulus of nanocomposites loaded with $\Phi \simeq 10$ vol.% in silica. On the other hand, polymers of different molecular weight resulted in shifted relaxation spectra,[3] leading therefore to an effective decorrelation consisting in a fixed $G'$ and a tunable $G''$. Note that although both grafted and free polymers in a given sample always had the same molecular weight (respectively denoted $N$ and $P$ in the literature), one could now imagine easily taking advantage of the $P/N$ ratio to further tune the nanocomposites structure and dynamics.[4, 5]

Beyond this somehow limited decorrelation of $G'$ and $G''$, intrinsically bounded by the Kramers-Kronig relationship, we have evidenced recently another remarkable decorrelation between linear and non-linear mechanics in similar materials.[6] This work was based on the surprising structural-insensitivity of the silica network to the addition of high fractions of short poly-propylene-glycol chains (up to 18 vol.%) acting as a covering agent. In agreement with our above-mentioned study, the identical filler dispersion was providing identical linear viscoelastic properties (and identical glass transition temperature),[2] regardless of the PPG layer's thickness for $\Phi$=15 vol.%. Nevertheless, the latter was found to play a major role at higher strain (from 1 to 100%) resulting in a pronounced softening of the nanocomposites both under oscillatory shear and steady tensile solicitations. While this result may very well open new perspectives to control materials' performances, it is also very likely to play a role in their processing route. Indeed, the control of the viscoelastic properties of crude rubber is a major issue for tires manufacturers. In this context, PPG could substitute some oils to act as a plasticizer during the process, without however deteriorating the properties of the materials under normal use conditions (strain < 100 % and frequency around 50 rad s$^{-1}$).

In this letter, we actually propose to go one step further by studying the vulcanized homologs of these SBR-PPG/Silica nanocomposites for which the composition is reported in Table 1. The components' list and their respective amount is usual for this kind of cross-linked materials, apart from the presence of PPG reported here for the first time to our knowledge. (Note that we provide an estimation of the PPG layer thickness, $\zeta$, assuming a total and homogeneous adsorption of the PPG chains on the nanoparticles of radius 13 nm).[6] For the sake of simplicity we make use of a single vulcanization accelerator (n-cyclohexyl-2-benzothiazyl-sulfenamide, denoted as CBS) while many other molecules among which TBBS,[7] TMTD[8] or DPG[8,9] are frequently used.

**Table 1**: Nanocomposites composition in phr (per hundred rubber). Note that the silica amount in phr is increasing while its volume fraction in the material is kept constant (15 vol.%). *PPG layer thickness (see text)

| (phr) | **15-0** | **15-0.5** | **15-1** | **15-3** | **15-5** | **15-8** |
|---|---|---|---|---|---|---|
| SBR | 100 | 100 | 100 | 100 | 100 | 100 |
| PPG | 0 | 1.3 | 2.7 | 8.5 | 15.2 | 27.1 |
| Silica | 41.2 | 41.8 | 42.4 | 45.2 | 48.3 | 53.9 |
| Sulfur | 2 | 2 | 2 | 2 | 2 | 2 |
| ZnO | 2 | 2 | 2 | 2 | 2 | 2 |
| Stearic acid | 1 | 1 | 1 | 1 | 1 | 1 |
| CBS | 2 | 2 | 2 | 2 | 2 | 2 |
| $\zeta$ (nm)* | 0 | 0.3 | 0.6 | 1.9 | 3.2 | 5.2 |

The first step of our investigation consists in verifying the isostructural character of our nanocomposites,[6] i.e., the insensitivity of the silica network to the presence of short PPG chains, once vulcanized. In Figure 1, we present transmission electron micrographs of the 15-0 and 15-8 vulcanized nanocomposites confirming similar structures in composites containing from 0 to 27 phr (18 vol.%) of PPG. Both materials are based on a network of fractal aggregates of ca. 100 nm of diameter, in conformity with our expectations for well-dispersed nanocomposites.[10] Note however that no SBR covalent grafting is used in the present study and that such good dispersions are believed to come from the mixing procedure we used – in particular from the shape of our "Roller" rotors.[6] In addition, the PPG adsorption onto the silica is clearly visible in the 15-8 sample, similarly as what we observed in crude materials.[6] In contrast with our previous study, no SAXS investigation was conducted here because of the presence of ZnO particles making any (presumably small) differences in scattering measurements, non-unequivocal. In fact, the SBR/ZnO X-ray contrast is estimated to be 12 times higher than its SBR/SiO$_2$ counterpart.

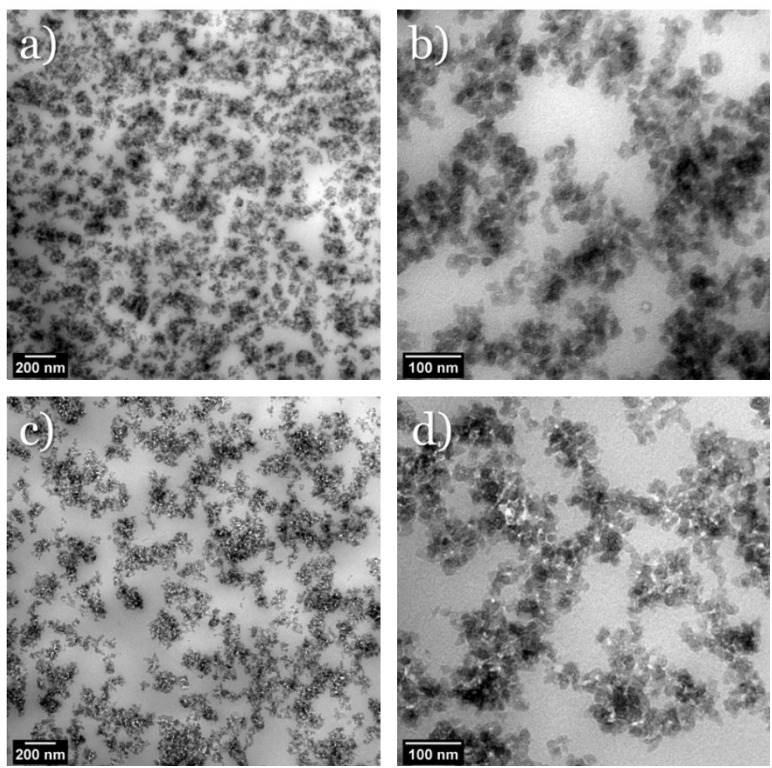

**Figure 1**: Transmission electron micrographs of nanocomposites. (a-b) 15-0 and (c-d) 15-8. Brightest zones in c-d stand for PPG adsorbed onto the silica surface.

Although the filler structure seems therefore not affected by the vulcanization process, the reciprocal is not true. In fact, nano-silica is expected to impact significantly the formation of the polymer network by interacting with the SBR chains (bound rubber) and most of the small polar molecules essential for the vulcanization.[11,12] Of a particular interest, Ramier et al. showed that the addition of silica in presence of CBS was accelerating significantly the vulcanization process.[9] They propose accordingly a probable adsorption of the accelerators at the silica surface, resulting into a heterogeneous vulcanization in the material – a high density of cross-links close to the silica particles and a looser polymer network far from them. Besides, they show in the same article that adding covering agents onto the nano-silica is deactivating this catalytic effect. In Figure 2, we present the normalized tensile storage modulus $E'(t)/E'_{end}$ measured as a function of time along with the vulcanization performed at 160 °C in our DMA apparatus for the whole set of samples. Figure 2a clearly evidences that adding silica accelerates the mechanical properties upturn with respect to the neat matrix (see $\tau_v^{SBR}$ and SI section 1). Then, by adding PPG, the vulcanization is unambiguously delayed, becoming even slower than in the neat SBR from 2.7 phr of PPG (15-1) and finally saturates at higher content because of mobility issues. The time corresponding to the inflexion points of such S-shaped curves ($\tau_v$) as well as the corresponding rate are reported in Figure 2b, illustrating the impact of PPG on the vulcanization kinetics.

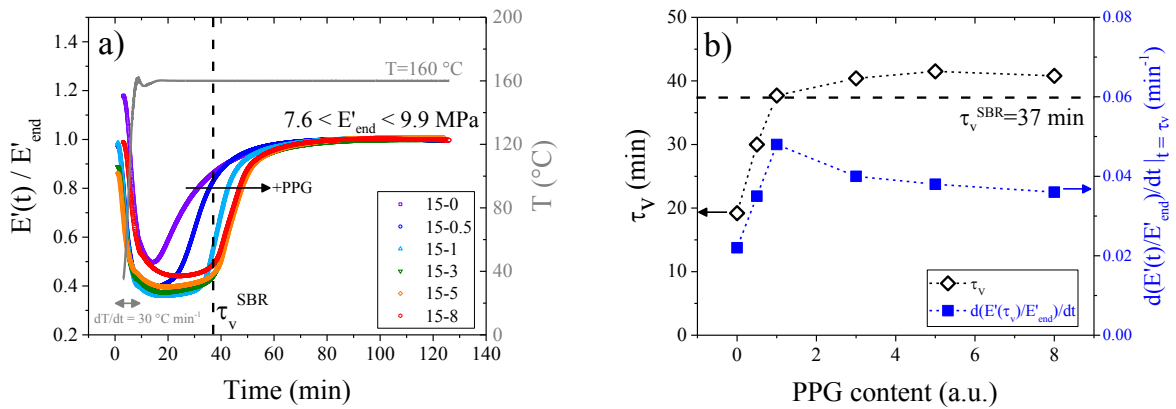

**Figure 2**: Vulcanization kinetics. a) $E'(t)/E'_{end}$ for nanocomposites and temperature as a function of time showing the rubber hardening at 160 °C. b) Characteristic time of the vulcanization process (inflexion point, $\tau_v$) and the corresponding rate as a function of the amount in PPG within the nanocomposite. The dashed lines in a-b stand for the characteristic vulcanization time of the neat SBR of corresponding formulation.

While the vulcanization kinetics are thus clearly perturbed by our covering agent, we also need to identify its effect on the resulting architecture of the polymer network. To access this information, we have combined both small amplitude oscillation shear (SAOS) and multiple-quantum nuclear magnetic resonance (MQ-NMR) experiments.

Similarly to the case of crude nanocomposites, Figure 3a reveals that the linear viscoelastic behavior of all the composites are quasi-identical. In fact, the ratio between the highest and lowest values of $G'$ and $G''$ are respectively 1.5 and 1.7, in line with the slight and erratic variations previously observed,[6] which corroborates the strong connection between the filler structure and the dynamic moduli. (Note that the reinforcement generated by 15 vol.% in silica is close to a factor 6 for both moduli, in agreement with a previous study.[13]) Interestingly, combining the identical viscoelastic properties with the identical silica structure in all the composites, suggests that the polymer network architecture must also be conserved in spite of different vulcanization kinetics. This latter aspect is investigated in Figure 3b where multiple quantum NMR intensity ($I_nDQ$) is compared for samples containing low amount of PPG and neat SBR (see additional data and discussion in SI section 2). This method, well described in ref.[14,15], is frequently used to probe the inhomogeneous polymer networks with the aim to extract cross-link density distributions in vulcanized rubbers.[16,17] Here, it clearly supports our rheological study by showing that the polymer network architecture remains unchanged with varying the silica and the PPG content. Fitting the neat SBR data with the "universal function" proposed in ref.[15] combined with a log-normal distribution of the dipolar coupling constant $D_{res}$ (eq.3 in ref.[18]) results in the distribution presented in the Figure 3b inset. The transformation of this data into a distribution of molecular weights between entanglements ($M_e$) can then be achieved through the empirical expression proposed by Mujtaba et al. for the SBR (Fig. 2b in ref.[16]) which provides us with $M_e^{NMR} = 1.8$ kg mol$^{-1}$. This value can be easily compared with our rheological data enabling the estimation of $M_e$ through $G_N = \rho RT/M_e^{Rheo}$, where $\rho$ is the SBR mass density, $R$ the perfect gas constant and $T$ the temperature, leading to $M_e^{Rheo}$ =2.35 kg mol$^{-1}$ (while it is reported to be $M_e$ =2.9 kg mol$^{-1}$ in uncrosslinked SBR[19]). The discrepancy between the two techniques likely originates from the value of the empirical coefficient (1.027 10$^{27}$ m$^{-3}$/kHz)[16] – related with the SBR chemical microstructure – enabling to connect $D_{med}$ and $M_e^{NMR}$. Besides, while $M_e^{NMR}$ was reported to decrease with the silica content in the literature,[16, 20] it is visibly not the case in the present work. While the reason for such difference is unclear to us, we believe that it comes from the different nature of our silica (Zeosil 1115 MP from Solvay, BEL) and vulcanization accelerator (CBS only).

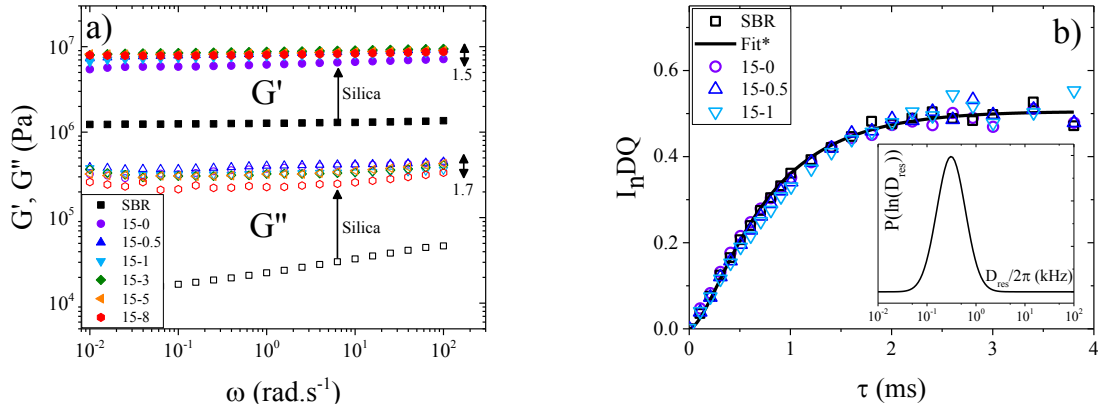

**Figure 3**: SBR and composites characterization at rest. a) Linear shear rheology. Storage ($G'$) and loss ($G''$) moduli as a function of the frequency for the whole set of nanocomposites. Double-arrows refer to the largest ratio observed between the maximal and minimal $G'$ and $G''$ when the composition is changed from 15-0 to 15-8. Temperature and strain amplitude are respectively fixed to 110 °C and 0.1 %. b) Low-field NMR $I_nDQ$ signal measured at 40 °C for neat SBR, and low PPG content samples. Solid line is fitted to the SBR data with the "universal function" (see text) resulting in $D_{med}/2\pi$ = 0.31 kHz and $\sigma_{ln}$ =0.67. Inset represents the corresponding $D_{res}$ log-normal distribution function.

Because they are characterized by similar filler structure and polymer network architecture, all the interest of the present rubbers resides therefore in their different behavior under large amplitude deformations. In our previous study, oscillatory strain sweeps performed up to $\gamma =100$ % revealed the softer nature of samples containing higher PPG content.[6] For the sake of comparison, we have repeated these experiments on their vulcanized homologs. The results are presented in Figure 4 showing a surprising insensivity of the mechanical behavior to the PPG content. In fact, both $G'(\gamma)$ (Figure 4a) and $G''(\gamma)$ (Figure 4b) appear to be quasi-identical for the whole series of composites.

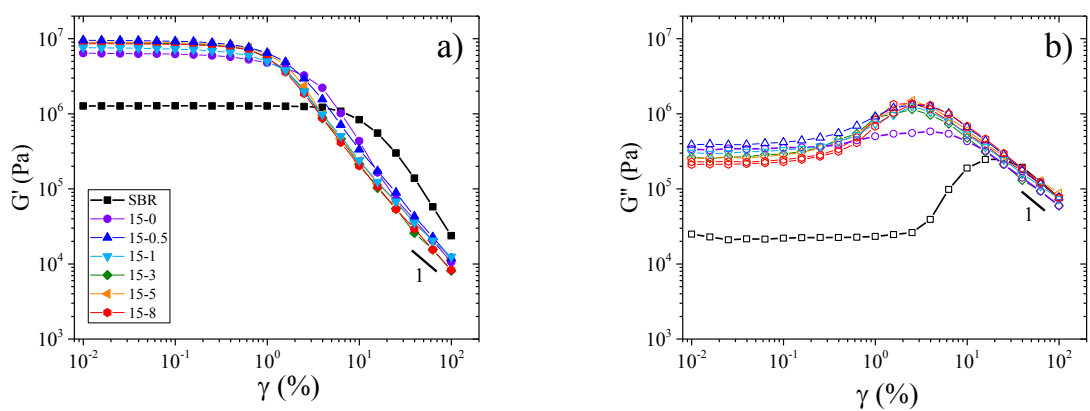

**Figure 4**: Dynamic strain sweeps performed at 110 °C to limit slip issues. a) Storage modulus ($G'$) as a function of the strain amplitude for the whole set of nanocomposites. The frequency is set to $\omega_0 =1$ rad s$^{-1}$. b) Analog data for the loss modulus ($G''$).

We believe that these observations must be understood in the following way (we refer the reader to Figure 6 for corresponding schematic representations):

(i) **At low strain amplitude ($\gamma$<1 %)**, the PPG at the silica surface does not impact the stress transmission through the vulcanized matrix nor the network made of silica aggregates and occluded rubber. (This is convincingly supported by Figure 3).

(ii) **For 1<$\gamma$<10 %**, the early drop of $G'$ in nanocomposites, known as the Payne effect, is believed to originate from the strain amplification within the polymer phase,[21] mostly dominated by the filler-filler interactions.[22] It seems insensitive to the presence of PPG supporting once again the isostructural character of our composites. The corresponding peak observed on $G''$ confirms the presence of a dissipative event, namely the failure of the filler network,[16] and the transition towards a pseudo-hydrodynamic regime. From a purely mechanical point of view, one can see this network failure as a parallel-to-series transition where the two components are (a) the reinforcing network made of silica and occluded-rubber, and (b) the softer vulcanized matrix.

(iii) The data at $\gamma > 10\ \%$, well supports this interpretation. While all the $G''$ overlap perfectly, confirming that the soft matrix dominates the viscoelastic response at high strain (and high strain rate since $\dot{\gamma}_{max} \sim \gamma$), $G'$ of all the composites fall below their SBR counterpart. Although it is unusual, we believe that this effect results from a severe strain amplification in the soft phase occurring in 15 vol.% filled materials, and that it mainly depends on the silica structure. In fact, a similar observation was made in our previous work on crude nanocomposites[6] indicating that the vulcanization process must not be responsible for it.

We then believe that the insensivity of the dynamic moduli to the PPG content can be rationalized through the filler mobility. As hypothesized in our previous work, we think that the PPG shell is playing the role of a lubricant which limits the long SBR chains adsorption[16,23] and facilitates both the rotational and translational motion of aggregates at long timescale and/or under large deformations. In other words, whereas aggregates could well diffuse in free SBR chains, we believe that additional crosslinks limit greatly their motion in vulcanized rubbers, including after the filler network failure ($10 < \gamma < 100\ \%$) – resulting in the disappearance of the progressive lubrication effect observed with increasing the PPG content (Fig. 6c-d in ref.[6]).

It seems therefore that small and medium strain amplitude solicitations are not sufficient to make emerge the different mechanical behavior of our composites in spite of their quite different formulations. In fact, we remind the reader that the 15-8 sample contains not less 18 vol.% in PPG, corresponding to 21 vol.% of the organic phase while the 15-0 is PPG free. We consequently present in Figure 5 more drastic tensile tests revealing finally a strong softening of the composites from $\varepsilon > 100\ \%$ with increasing the PPG content. In addition to the pronounced lowering of the $\sigma(\varepsilon)$ curves (Figure 5a), this softening is visible from the drop of the modulus at failure ($E_{max}$, Figure 5a-inset) as well as the level of the corresponding maximal stress ($\sigma_{max}$, Figure 5b). Besides, these experiments reveal a quasi PPG-independent strain at rupture ($\varepsilon_{max}$, Figure 5b), corroborating the isostructural character of the polymer matrix in all the nanocomposites.

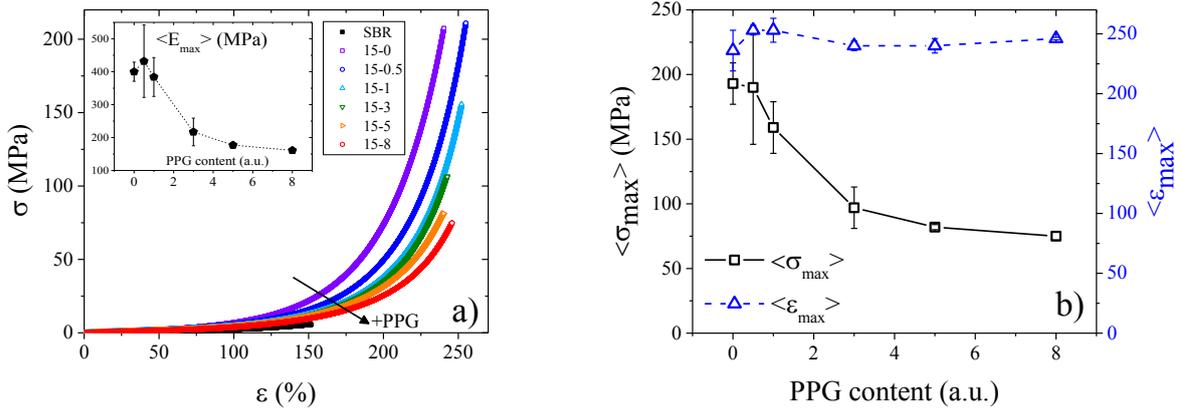

**Figure 5**: Tensile tests performed at room temperature. a) True stress as a function of the true longitudinal deformation for the whole set of nanocomposites and the neat vulcanized SBR. Inset stands for the apparent modulus at failure as a function of the PPG content. b) Maximal stress and strain at rupture as a function of the PPG content.

For the sake of clarity, we provide in Figure 6 our structural interpretation of the different mechanical behaviors involved in 15-0 and 15-8 materials from low to high strain amplitude, based on the results from Figure 4 and 5. The main message can be summarized in three points:

(i) The reinforcement at low strain amplitude is driven by the rigid network only, i.e., by the silica structure and the resulting occluded rubber existing between the aggregates. This rigid network and the vulcanized matrix works in parallel, insuring a high elasticity ($G' \approx$ 10 MPa), regardless of the PPG amount adsorbed at the interface.

(ii) Between 1 and 10% in deformation, the rigid network fails. Slight relative motions of the aggregates provoke the rupture of the occluded rubber bridges resulting in a strong energy dissipation. Here again, the PPG adsorbed at the interface (without covering it fully) plays no significant role.

(iii) At higher deformation, the aggregates are not connected anymore (pseudo-hydrodynamic regime). Their rotational and translational diffusion is enhanced by the presence of PPG (lubrication), favoring the materials relaxation, i.e., lower mechanical properties. The contribution of the rigid network progressively vanishes with the strain amplitude and the PPG content, making the vulcanized matrix the dominant elastic component – see notably the neat SBR vs. 15-8 tensile curves being almost overlapped around 150 % in deformation.

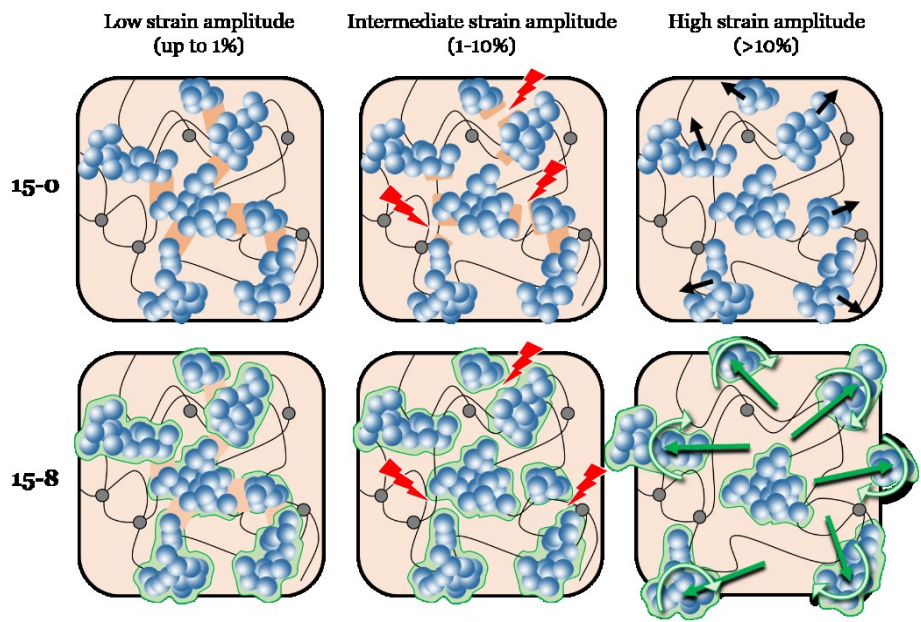

**Figure 6**: Schematic representation of the nanocomposites structure at different strain amplitude for the two extreme composition: 15-0 and 15-8. Blue beads stand for the silica, black lines and grey dots for the vulcanized matrix. Orange and green zones represent the occluded rubber and the adsorbed PPG respectively. Red "lightning" represent the network failure. Black and green arrows represent the aggregates' motion in 15-0 and 15-8 respectively (translation and rotation).

In conclusion, we have evidenced that the isostructural softening effect caused by the addition of short chains of PPG within SBR/silica crude composites was preserved into their vulcanized counterparts. Remarkably, we have demonstrated that both the filler structure and the cross-link density distributions of composites containing from 0 to 18 vol.% in PPG was unchanged in spite of quite different vulcanization kinetics. This result was further corroborated by small and medium amplitude oscillation shear experiments showing similar viscoelastic responses in all the materials conversely to what was observed in crude nanocomposites. This qualitative difference is supporting our hypothesis on the lubricant role of the PPG confirming the importance of the filler dynamics in nanocomposites. Large amplitude deformations make finally appear the PPG signature by revealing a pronounced softening of the composites enabling somehow the decorrelation between the normal use of such operative materials (strain < 100 % and $\dot{\gamma} \approx 50$ rad s$^{-1}$) and their urgent needs to dissipate energy (strain > 100 % and $\dot{\gamma} > 10$ krad s$^{-1}$).

**Experimental Section**

**Samples preparation**. SBR/silica (Zeosil 1115MP) nanocomposites containing various fractions of PPG (4 kg mol$^{-1}$) were processed through internal mixing. The glass transition temperature of the SBR is -37 °C, its molecular weight 267 kg mol$^{-1}$ and its polymolecularity index is 1.09. Additional details on the materials and methods are reported in our previous study on unvulcanized nanocomposites.[6] Vulcanization ingredients ((i) stearic acid, (ii) ZnO, (iii) sulfur and CBS) were then incorporated stepwise to the materials through 12 successive two-rolls milling performed at 80 °C. After every passage, the flat sheet of nanocomposite was rolled up to form a "cigar" and inserted perpendicularly to the rolls axes. The vulcanization was finally carried out in a hot-press at 160 °C and 25 MPa for 80 minutes, providing 1 mm thick samples usable for the different characterizations. The corresponding heating rate was measured through the incorporation of a thermocouple within the materials resulting in ca. 30 °C min$^{-1}$. This value was set in our DMA experiments (Fig. 2a) to follow the evolution of the mechanical properties along with the vulcanization in similar conditions.

**MQ-NMR.** The 1H MQ solid-state NMR experiments were performed on a Bruker minispec mq20 spectrometer ("NF" electronics) operating at a resonance frequency of 20 MHz with 90° pulse length of 3.1 µs and a dead time of 15 µs. The experiments and the analysis of the measured raw data were performed following the previously published procedures,[14] including the subtraction of signal tails related to network defects prior to calculating $I_nDQ$.

**Transmission electron microscopy and mechanical analysis** (SAOS, strain sweeps and tensile tests) – see ref.[6] for details, present procedures are identical.

## Associated content

## Author Information

Corresponding author: Guilhem P. Baeza, guilhem.baeza@insa-lyon.fr

ORCID:

Guilhem P. Baeza - 0000-0002-5142-9670

Florent Dalmas - 0000-0002-6179-0881

Fabien Dutertre - 0000-0001-5818-7902

Notes: The authors declare no competing financial interest.

## Acknowledgements

G.P.B. and F.D. are grateful Hocine Houat (INSA-Lyon) for his help with the samples preparation. G.P.B. warmly thank Kay Saalwächter (Univ. Halle) and Walter Chassé (Univ. of Münster) for enlightening discussions about the low-field NMR experiments and corresponding data treatment. All the authors thank Jean-Marc Chenal and Laurent Chazeau (INSA-Lyon) for sharing their experience on vulcanization. All the authors are indebted to the "Microstructure Technological Center" (CTµ) of University of Lyon for the access to the transmission electron microscope and the ultramicrotomy device.

# References

1. Ahmad, S.; Schaefer, R. J., Energy saving tire with silica-rich tread. Google Patents: 1985;

2. Baeza, G. P.; Genix, A.-C.; Degrandcourt, C.; Gummel, J. r. m.; Mujtaba, A.; Saalwächter, K.; Thurn-Albrecht, T.; Couty, M.; Oberdisse, J. Studying Twin Samples Provides Evidence for a Unique Structure-Determining Parameter in Simplifed Industrial Nanocomposites. *ACS Macro Lett.* **2014,** *3*, 448-452

3. Baeza, G. P.; Genix, A.-C.; Degrandcourt, C.; Petitjean, L.; Gummel, J.; Schweins, R.; Couty, M.; Oberdisse, J. Effect of Grafting on Rheology and Structure of a Simplified Industrial Nanocomposite Silica/SBR. *Macromolecules* **2013,** *46*, 6621-6633

4. Kumar, S. K.; Benicewicz, B. C.; Vaia, R. A.; Winey, K. I. 50th Anniversary Perspective: Are Polymer Nanocomposites Practical for Applications? *Macromolecules* **2017,** *50*, 714-731

5. Sunday, D.; Ilavsky, J.; Green, D. L. A phase diagram for polymer-grafted nanoparticles in homopolymer matrices. *Macromolecules* **2012,** *45*, 4007-4011

6. Trinh, G. H.; Desloir, M.; Dutertre, F.; Majesté, J.-C.; Dalmas, F.; Baeza, G. P. Isostructural softening of the filler network in SBR/silica nanocomposites. *Soft Matter* **2019**,

7. Hosseini, S. M.; Razzaghi-Kashani, M. Catalytic and networking effects of carbon black on the kinetics and conversion of sulfur vulcanization in styrene butadiene rubber. *Soft Matter* **2018,** *14*, 9194-9208

8. Ohnuki, T. The vulcanizing system of diene rubber. *Int. Polym. Sci. Techol.* **2015,** *42*, 39-46

9. Ramier, J.; Chazeau, L.; Gauthier, C.; Guy, L.; Bouchereau, M.-N. Influence of silica and its different surface treatments on the vulcanization process of silica filled SBR. *Rubber Chem. Technol.* **2007,** *80*, 183-193

10. Genix, A.-C.; Baeza, G.; Oberdisse, J. Recent advances in structural and dynamical properties of simplified industrial nanocomposites. *Eur. Polym. J.* **2016,** *85*, 605-619

11. Song, L.; Li, Z.; Chen, L.; Zhou, H.; Lu, A.; Li, L. The effect of bound rubber on vulcanization kinetics in silica filled silicone rubber. *RSC Adv.* **2016,** *6*, 101470-101476

12. Hosseini, S. M.; Razzaghi-Kashani, M. Vulcanization kinetics of nano-silica filled styrene butadiene rubber. *Polymer* **2014,** *55*, 6426-6434

13. Baeza, G. P.; Genix, A.-C.; Degrandcourt, C.; Petitjean, L.; Gummel, J.; Couty, M.; Oberdisse, J. Multiscale Filler Structure in Simplified Industrial Nanocomposite Silica/SBR Systems Studied by SAXS and TEM. *Macromolecules* **2013,** *46*, 317-329

14. Saalwächter, K. Proton multiple-quantum NMR for the study of chain dynamics and structural constraints in polymeric soft materials. *Prog. Nucl. Magn. Rson. Spectrosc.* **2007,** *51*, 1-35

15. Chassé, W.; Valentín, J. L.; Genesky, G. D.; Cohen, C.; Saalwächter, K. Precise dipolar coupling constant distribution analysis in proton multiple-quantum NMR of elastomers. *J. Chem. Phys.* **2011,** *134*, 044907

16. Mujtaba, A.; Keller, M.; Ilisch, S.; Radusch, H. J.; Thurn-Albrecht, T.; Saalwächter, K.; Beiner, M. Mechanical Properties and Cross-Link Density of Styrene–Butadiene Model Composites Containing Fillers with Bimodal Particle Size Distribution. *Macromolecules* **2012,** *45*, 6504-6515

17. Grasland, F.; Chazeau, L.; Chenal, J.-M.; Caillard, J.; Schach, R. About the elongation at break of unfilled natural rubber elastomers. *Polymer* **2019,** *169*, 195-206

18. Jakisch, L.; Garaleh, M.; Schäfer, M.; Mordvinkin, A.; Saalwächter, K.; Böhme, F. Synthesis and structural NMR characterization of novel PPG/PCL conetworks based upon heterocomplementary coupling reactions. *Macromol. Chem. Phys.* **2018,** *219*, 1700327

19. Fetters, L. J.; Lohse, D. J.; Colby, R. H., Chain Dimensions and Entanglement Spacings. In *Physical Properties of Polymers Handbook*, Mark, J. E., Ed. Springer New York: New York, NY, 2007; pp 447-454

20. Sotta, P.; Albouy, P.-A.; Abou Taha, M.; Long, D. R.; Grau, P.; Fayolle, C.; Papon, A. l. Nonentropic Reinforcement in Elastomer Nanocomposites. *Macromolecules* **2017,** *50*, 6314-6322

21. Domurath, J.; Saphiannikova, M.; Ausias, G.; Heinrich, G. Modelling of stress and strain amplification effects in filled polymer melts. *J. Non-Newtonian Fluid Mech.* **2012,** *171*, 8-16

22. Warasitthinon, N.; Genix, A.-C.; Sztucki, M.; Oberdisse, J.; Robertson, C. G. The Payne Effect: Primarily Polymer-Related or Filler-Related Pehnomenon? *Rubber Chem. Technol.* **2019**, *92*, 599-611

23. Baeza, G. P.; Oberdisse, J.; Alegria, A.; Saalwächter, K.; Couty, M.; Genix, A.-C. Depercolation of aggregates upon polymer grafting in simplified industrial nanocomposites studied with dielectric spectroscopy. *Polymer* **2015,** *73*, 131-138